\documentclass[useAMS]{mn2e}
\usepackage{graphicx,graphics,times}
\def\spose#1{\hbox to 0pt{#1\hss}}
\def\mdot {\spose{\raise 7.0pt\hbox{\hskip 5.0pt{\char '056}}}M}
\newcommand{\kms}{km\,s$^{-1}$}
\newcommand{\msol}{M$_\odot$}

\newcommand{\rsol}{R$_\odot$}

\begin{document}

\title
[Radiative-transfer models of WR104]
{Three-dimensional dust radiative-transfer models: \\The Pinwheel Nebula
  of WR104}
\author[Tim J. Harries et al.]{Tim J. Harries$^1$, John D. Monnier$^2$, Neil H. Symington$^1$, and
  Ryuichi Kurosawa$^1$ \\
$^1$School of Physics, University of Exeter, Stocker Road, Exeter EX4
4QL, England. \\
$^2$University of Michigan, 941 Dennison Building, 500 Church Street, Ann
Arbor, MI 48109-1090, U.S.A. \\
}
\date{Dates to be inserted}

\maketitle

\begin{abstract}
  
  We present radiative-transfer modelling of the dusty spiral Pinwheel
  Nebula observed around the Wolf-Rayet/OB-star binary WR104. The
  models are based on the three-dimensional radiative-transfer code
  {\sc torus}, modified to include an adaptive mesh
  that allows us to adequately resolve both the inner spiral turns
  (sub-AU scales) and the outer regions of the nebula (distances of
  10$^4$ AU from the central source).  The spiral model provides a
  good fit to both the spectral energy distribution and Keck
  aperture masking interferometry, reproducing both the maximum
  entropy recovered images and the visibility curves. We deduce a dust
  creation rate of $8\pm 1 \times 10^{-7}$ M$_\odot$\,yr$^{-1}$,
  corresponding to approximately 2\% by mass of the carbon produced by
  the Wolf-Rayet star. Simultaneous modelling of the imaging and
  spectral data enables us to constrain both the opening-angle of the
  wind-wind collision interface and the dust grain size.  We conclude
  that the dust grains in the inner part of the Pinwheel nebula are
  small ($\sim100$\AA), in agreement with theoretical
  predictions, although we cannot rule out the presence of larger
  grains ($\sim 1$\micron) further from the central binary. The
  opening angle of the wind-wind collision interface appears to be
  about $40^\circ$, in broad agreement with the wind parameters
  estimated for the central binary. We discuss the success and
  deficiencies of the model, and the likely benefits of applying
  similar techniques to the more the more complex nebulae observed
  around other WR/O star binaries.

\end{abstract}


\section{Introduction}

Wolf-Rayet stars are the descendants of massive O-stars whose spectra
are characterized by broad emission lines that are formed in their
dense, high-speed stellar winds. Wolf-Rayet (WR) stars may be
classified, by their spectral features, into nitrogen-rich (WN) and
carbon-rich (WC) sequences. The current status of research into WR
stars in our own and starburst galaxies is reviewed by Conti (2000).

The coolest WC stars often have an excess of IR radiation, indicative
of warm circumstellar dust (Allen, Harvey \& Swings 1972; Cohen,
Barlow \& Kuhi 1975; Williams, van der Hucht \& Th\'{e} 1987). Systems
with a near-constant excess are termed persistent dust producers,
while those that show periodic behaviour in their IR fluxes are
designated `episodic' (e.g. Williams 1995). The fact that dust can
condense and survive in the harsh UV-radiation field close to a WR
star is somewhat surprising, and is a challenging problem for
astrochemistry. The presence of dust in a highly-ionized environment
also has ramifications for those studying star formation rates (SFR),
where the presence of dust hinders the accurate calibration of SFR
diagnostics such as H$\alpha$ luminosity (see, e.g, Dopita et
al. 2003).

The link between binarity and dust formation in WC stars was forged
with observations of WR140, which demonstrated that the dust
production episodes coincided with periastron (see, e.g. Williams
1999). It appears that the higher density associated with the
wind-wind collision region aids dust condensation, and this
interpretation was strengthened by high-resolution imaging of the new
dust produced during periastron (Monnier et al. 2002). Further links
between the episodic dust production and binarity have been
established for WR125 (Williams et al. 1994) and WR137 (Williams et
al. 1985), while significant progress has been made in understanding
the conditions in the innermost regions of colliding wind binaries by
combining hydrodynamical modelling with radiative-transfer modelling
in the radio regime (Dougherty et al.  2003).

Persistent dust makers {\em may} be associated with single stars.
Zubko (1998) performed a sophisticated calculation of the formation of
dust in WR stars, adopting a spherical dust shell composed of
amorphous carbon grains.  The dynamics and grain growth/destruction
mechanisms were accounted for, and it was found that the grains grew
to 100--200\AA\ as a result of implantation of carbon ions.
Approximately 1\% of the carbon emitted by the WR star was found to
condense into dust. The rate of grain seed production was determined
for individual systems by fitting the observed spectral energy
distributions (SEDs). Although the Zubko calculations indicate that
grain growth can occur in single systems, Cherchneff et al. (2000)
suggested that significantly higher densities than those found in a
smooth wind were required to produce the dust precursors. In fact, it
has long been suggested that the persistent dust makers are binary
system whose separation is such that the critical density for dust
production is always achieved in the wind-wind interface (Williams \&
van der Hucht 1992).

If the formation of dust in WC systems does occur in the wind-wind
collision interface, then it appears unlikely that the dust will be
spherically distributed, and high resolution IR imaging seems to
support this.  Marchenko, Moffat \& Grosdidier (1999) used maximum
entropy reconstruction of NICMOS-2 imaging of WR137 to reveal an
elongated structure.  High-resolution imaging of WR104 (Tuthill,
Monnier \& Danchi 1999), obtained using Keck aperture masking
interferometry, revealed the presence of a spiral distribution of
dust, viewed at low inclination (11$^\circ$).  Multi-epoch
observations demonstrated that the spiral rotates with a period of
245d. The linear archimedian nature of the spiral suggests that it is
associated with a flow of material away from a binary. Lunar
occultation observations in the $K$-band confirm the spiral structure
(Mondal \& Chandrasekhar 2002).  A second `pinwheel nebula' has been
observed in WR98a (Monnier, Tuthill \& Danchi 1999), while a
structured nebula, of larger extent, was observed in WR112 using
imaging at 18.2\micron\ by Marchenko et al. (2002), although its
pinwheel shape is more poorly defined.

In order to understand the formation mechanism of dust in colliding
wind systems we must obtain more information on the circumstellar
environment such as its geometric properties, its temperature, and its
dust grain size distribution. In this paper we describe
three-dimensional radiative transfer models of the extended dusty
nebula around WR104, using a state-of-the-art code. In the next
section we introduce the modelling code, and present tests of the code
against an established one-dimensional code and a two-dimensional
benchmark. We then describe the adopted model parameters, including
geometry, the SEDs of the binary components, and the grain composition
and size distribution. In subsequent sections we compare the resulting
images and optical/infrared SED with the WR104 observations, and
discuss the constraints the models place on the dust formation
mechanisms.

\section{The {\sc torus} radiative-transfer code}

The radiative-transfer modelling was performed using the {\sc torus}
code (Harries 2000). Originally written to model line-transfer in a
moving medium, the code is based on Monte-Carlo (MC) techniques and
tracks photons in three-dimensions. Variance reduction techniques are
used to minimize the sampling errors that are always associated with
MC methods. Dust scattering physics had already been included in {\sc
  torus} (see Harries, Babler, \& Fox 1999), but to model the IR
continuum it was necessary to introduce dust radiative equilibrium
physics. We additionally incorporated an adaptive mesh gridding, which
allows us to easily treat the large range of scales in the problem.

\subsection{Radiative equilibrium}

Traditional radiative-equilibrium calculations rely on finding a
numerical solution of the equation of radiative transfer. This is a
relatively straight-forward method, but is typically limited to
geometries with spherical (1-d) or rotational (2-d) symmetry (although
see Steinacker et al. 2003).

Lucy's (1999) MC-based method allows one to perform a full
radiative-equilibrium calculation on an arbitrary dust distribution.
Photon `packets' are produced at the stellar photosphere and followed
through the grid, being scattered, or absorbed and re-emitted, until
they eventually escape the computational domain or are re-absorbed by
the stellar photosphere. An extensive description of the algorithm is
given by Lucy (1999) and Kurosawa et al. (2004), and we note that we
have adopted Lucy's method without modification.

\subsection{Adaptive mesh}

Complex three-dimensional structures may not be properly spatially
sampled using traditional Cartesian or spherical polar grids, which
often do not make efficient use of the available computer memory. A
better method of producing a spatial grid is to employ a mesh that has
higher spatial resolution in the regions of interest (high density or
high opacity gradient) and a lower resolution in others (e.g. Kurosawa
\& Hiller 2001; Steinacker et al. 2003). 

In {\sc torus} we have implemented an adaptive mesh consisting of
nested cubes, with a factor two in linear scale between each nesting
level. In other words, each parent cube may be split into 8 children,
each of which may in turn have 8 children of their own and so on. The
grid data are stored as on octal tree, which is manipulated via
recursive algorithms. The grid refinement can occur according to a
variety of different criteria. For example, if there is an analytical
formula for the density, the cells can be recursively split until each
cell has a mass no greater than a pre-defined limit, or until no cell
has an optical depth greater than a certain value. (The latter
algorithm was used for the 2-D benchmark model). The {\sc torus}
implementation of the adaptive mesh refinement algorithm is described
in detail by Symington (2004).

For the case under consideration here we take a list of particle
positions (produced by a ballistic particle model) and refine the
grid until a cell contains no more than a certain number of
particles. The relative density of the cell is then simply given by its
volume and the number of particles it contains. The density of the
computational domain is then integrated and rescaled to match the
required envelope mass.

\subsection{Code testing}

Many tests were run to ensure that the new implementation of the
algorithm reproduced known solutions. The simplest tests involved
checking that energy is conserved by the model. We then benchmarked
the code against the established 1-d dust transfer code {\sc dusty}
(Ivezic, Nenkova \& Elitzur 1999).

\begin{figure*}
\includegraphics[width=180mm]{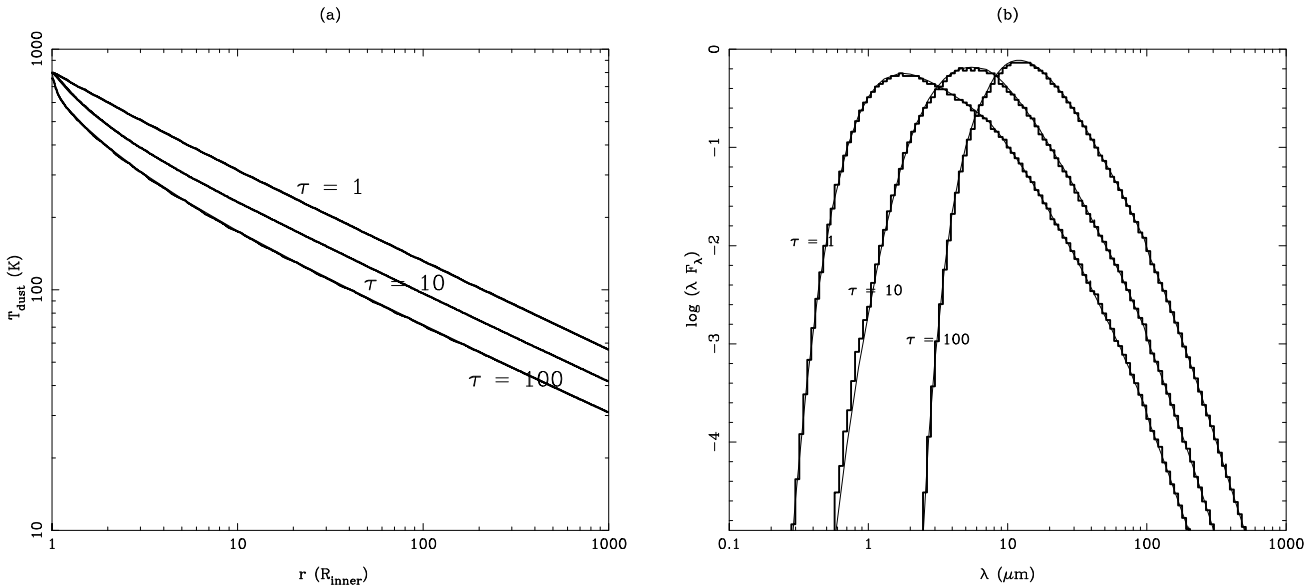}
\caption{A comparison between dust radiative equilibrium models
  computed by {\sc dusty} (continuous lines) and {\sc torus}
  (histogram). Panel (a) shows the radial temperature run, while (b)
  shows the emergent SED.}
\label{dusty_fig}
\end{figure*}

For the test model we adopted a thick spherical shell ($r_{\rm
  outer}=1000r_{\rm inner}$) with $\rho(r) \propto r^{-2}$, composed
of amorphous carbon dust with an MRN (Mathis, Rumpl, \& Nordsieck
1977) size distribution, and with the temperature of the inner edge of
the shell fixed at 800\,K. Three models were computed, corresponding
to optical depths at 5500\AA ($\tau_{5500}$) of 1, 10, and 100.  We
found excellent agreement between the radial dust temperature run
computed by {\sc torus} and {\sc dusty}, and the model SEDs show
similarly excellent agreement (see Figure~\ref{dusty_fig}).

\begin{figure*}
\includegraphics[width=180mm]{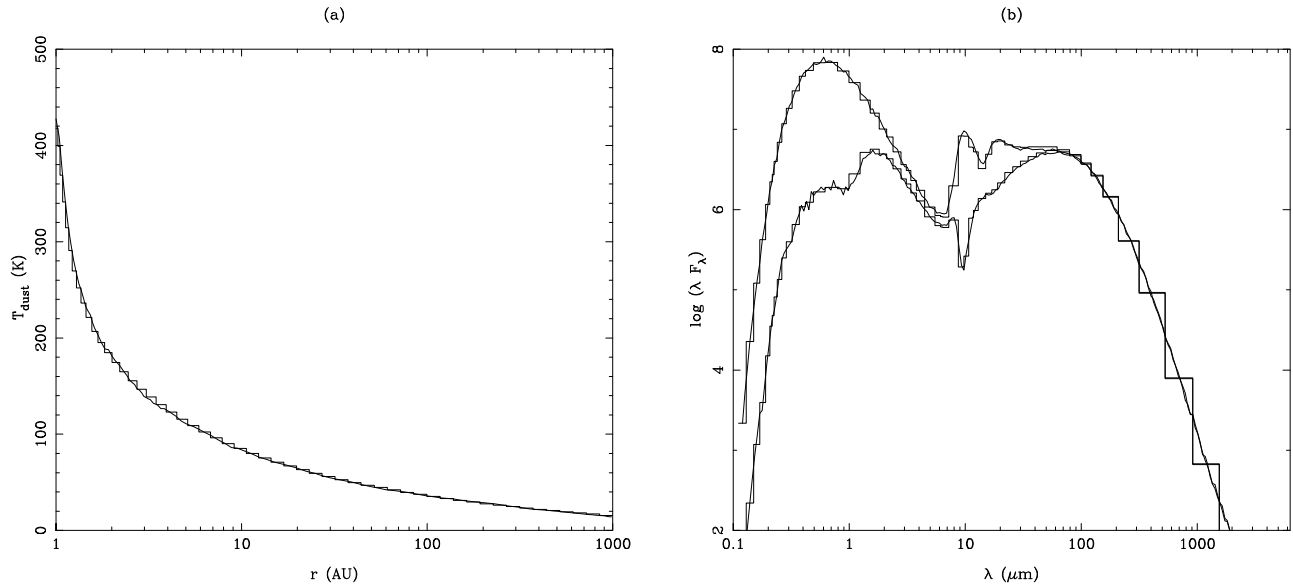}
\caption{Comparison between the optically thick ($\tau_{5500} = 100$) 
  2-D dusty disc benchmark model of Pascucci et al. (2003) and {\sc
    torus}. Figure (a) shows the radial run of temperature through the
  disc at an angle of $2.5^\circ$ to the equator for {\sc torus}
  (continuous line) and {\sc MC3D} (histogram). Figure (b) shows the
  emergent SED for the same model viewed at inclinations of
  $12.5^\circ$ (top line) and $77.5^\circ$ (lower line).}
\label{benchmark_fig}
\end{figure*}

We subsequently extended our tests to two dimensions, using the dusty
disc benchmark proposed by Pascucci et al. (2003), which in turn is
based on the disc geometry of Shakura \& Sunyaev (1973). Comparison
between the radial and vertical temperature structure showed excellent
agreement with the benchmark results, even for the most optically
thick ($\tau_{5500}=100$) case (Figure~\ref{benchmark_fig}a). The SED
computed using {\sc torus} also shows good agreement with the
benchmark results (Figure~\ref{benchmark_fig}b). 

\section{WR104 observations}

Previous models of dust formation in WR104 have relied solely on
fitting the SED, and so inevitably have been based on the simplest
plausible geometry, that of a thick spherical shell. In this paper we
employ the additional observational constraint provided by
ultra-high-resolution imaging.

\subsection{The spectral energy distribution}

The longest wavelengths in the SED of WR104 are defined by IRAS
observations at 12, 25 and 60 microns (Cohen 1995).  Overlapping with
this photometry is an ISO-SWS spectrum covering the wavelength range
2.4--40 microns (spectral resolution 250--600) described by van der
Hucht et al. (1996) and obtained from the ISO archive. Our JHK
photometry is taken from Williams, van der Hucht \& Th\'{e} (1987),
along with their $V$-band value and the $B$-band spectrophotometry of
Torres-Dodgen \& Massey (1988).

We note that the short-wavelength region of the SED should be treated
with caution, since WR104 shows significant ($>1^{\rm m}$) variability
in the optical (Crowther 1997; Kato et al. 2002), and unfortunately
the optical, near-IR, and mid-IR data are not contemporaneous.

\subsection{High resolution imaging}

Aperture masking interferometry is performed by placing aluminium masks
in front of the Keck-I infrared secondary mirror.  This technique
converts the primary mirror into a VLA-style interferometric array,
allowing the Fourier amplitudes and closure phases for a range of
baselines to be recovered with minimal `redundancy' noise e.g. Baldwin
et al. (1986), Jennison (1958).  The Maximum Entropy Method (MEM)
(Skilling \& Bryan 1984, Narayan \& Nityananda 1986) has been used to
reconstruct diffraction-limited images from the interferometric data,
as implemented in the VLBMEM package by Sivia (1987).  In order to
check the reliability of the reconstructions, the MEM results have
been compared with those from the CLEAN reconstruction algorithm
(H\"{o}gbom 1974, Cornwell \& Wilkinson 1981, Pearson \& Readhead
1984).  Further engineering and performance details may be found in
Tuthill et al. (2000) and Monnier (1999).

WR~104 was observed on U.T. 1999 April 15 and 1999 June 5 Keck-I using
the Near Infrared-Camera (Matthews \& Soifer 1994, Matthews et al.
1996) in speckle mode with an integration time of 0.137\,s per frame,
employing a circular `annulus' mask.  We present here imaging
results using two bandpasses: H filter ($\lambda_0 = 1.6575$\,$\mu$m,
$\Delta\lambda = 0.333$\,$\mu$m) and the CH$_4$ filter ($\lambda_0 =
2.269$\,$\mu$m, $\Delta\lambda = 0.155$\,$\mu$m).  The observations of
WR~104 modelled in this paper were first published in Tuthill et al.
(1999) and additional details can be found there.

\section{Model parameters}

The parameters of our model may be divided into `geometrical' and
`physical'. The geometrical parameters describe the distribution of
dust around the binary in three dimensions. The physical parameters
describe the properties of the stars themselves, as well as the dust
grain size and chemistry.

\subsection{Physical parameters}

It appears that the binary system is composed of a WC9 star and an
late-O or early-B dwarf companion, with a luminosity ratio (WC:OB) of
1:2 (Crowther 1997). Using $\log(L/L_\odot)=4.6$ for the WC component
(Crowther 1997), and adopting an effective temperature of 30\,kK for
the OB star, we used the luminosity ratio to determine the radius of
the OB component of 10\,\rsol.  The input SED is a Kurucz Atlas9 model
atmosphere ($\log g=4$) for the OB star, and a CMFGEN (Hillier \& Lanz
2001) model for the WC star (Crowther, priv.  comm.).

\begin{figure}
\includegraphics{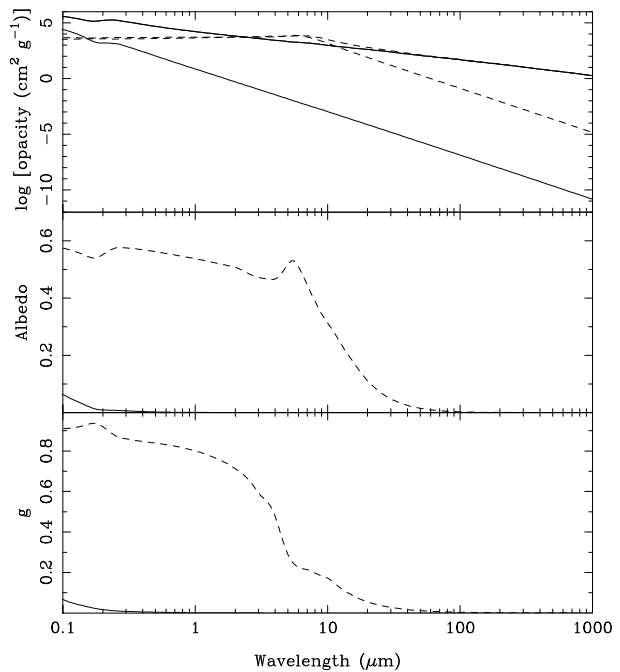}
\caption{Properties of the amorphous carbon grains as a function of
  wavelength. The top panel shows the absorption (top solid line) and
  scattering opacities (bottom solid line) of the 0.1\micron\ grains,
  while the middle panel shows the albedo. The bottom panel shows the
  standard Henyey-Greenstein $g$-factor (but note that we use the full
  Mie-scattering phase matrix in our calculations). Also plotted are
  the equivalent curves (dashed lines) for the 1\micron\ grains
  discussed in Section~\protect\ref{grainsize_sec}.}
\label{albedo_fig}
\end{figure}

We take our dust parameters from Zubko (1998), who found that
amorphous carbon dust grains may form rapidly from small (5\AA) seeds
to small grains (100\AA) by the implantation of impinging carbon ions.
Since the grain growth occurs relatively quickly, we adopt a uniform
grain radius of 100\AA\ (alternative grain size distributions are
considered in Section~\ref{grainsize_sec}). The optical constants for
the grains were taken from the pure amorphous carbon sample from Zubko
et al. (1996), and the opacity and albedo of the grains are shown as a
function of wavelength in Figure~\ref{albedo_fig}.

\subsection{Geometrical parameters}

The geometrical parameters of our model are determined from the
interferometric imaging. The simplest model that fits the data is that
of a linear spiral, viewed at an inclination of 11$^\circ$ (Tuthill et
al. 1999). The radial motion of the spiral, determined from the 245\,d
period is 0.303~mas/yr (Tuthill et al. 1999). The fact that we measure
the angular speed of the outflow leads to a degeneracy between outflow
speed and the systemic distance.  For our reference model we adopt the
canonical distance of 1.6\,kpc (e.g. Lundstrom \& Stenholm 1984),
which implies an outflow speed of 845\,km\,s$^{-1}$. We note that this
is the reverse approach to that adopted by Tuthill et al. (1999), who
used the WR terminal velocity (1220 \kms, Howarth \& Schmutz 1992) to
constrain the distance. A possible problem the latter method is that
the WR terminal velocity is obtained from the emission line
morphology, which is formed in the wind before the wind-wind collision
shock. This velocity may not necessarily be representive of the
post-shock flow, if for example the WR wind is subjected to radiative
braking (e.g. Gayley, Owocki \& Cranmer 1997) near the O~star.

The three dimensional structure of the dusty spiral is thought to stem
from a conic ejection of material from the wind-wind collision
interface. The precise geometry of this surface is unknown, although
analytical approximations to its shape may be determined from
momentum-balance arguments. Eichler \& Usov (1993) give the following
formula for the semi-opening angle ($\theta$, in radians)
\begin{equation}
\theta = 2.1 \left( 1- \frac{\eta^{2/5}}{4} \right) \eta^{1/3}
\label{theta_eq}
\end{equation}
where
\begin{equation}
\eta = \frac{\mdot_{\rm OB} \, v_{\infty, {\rm OB}}}{\mdot_{\rm WR} \,
  v_{\infty, {\rm WR}}}.
\end{equation}
For the WR component we use a mass-loss rate of $3 \times
10^{-5}$\,\msol\,yr$^{-1}$ (Crowther 1997) and a (pre-shock) terminal
velocity of 1220 \kms (Howarth \& Schmutz 1992). For the O-star
component we use the empirical mass-loss rate/luminosity relation of
Howarth \& Prinja (1989), which gives $6 \times
10^{-8}$\,\msol\,yr$^{-1}$, and adopt a plausible value of 2000\,\kms\ 
for the terminal velocity. These quantities give $\theta \approx
20^\circ$, and for our reference model we assume that the outflow is
spread over small range of semi-opening angles of 17.5--22.5$^\circ$.
We investigate the effect of changing the cone angle in
Section~\ref{angle_sec}.

\begin{figure}
\includegraphics[width=88mm]{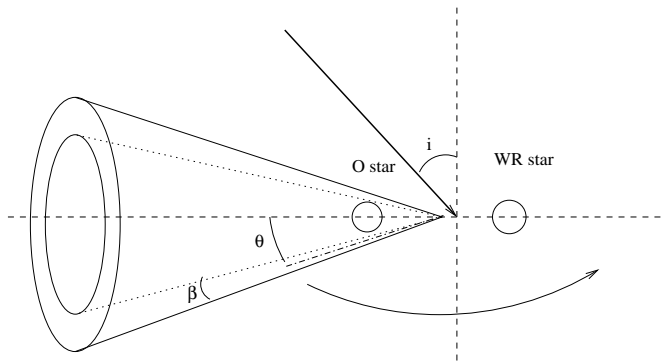}
\caption{The conic geometry used for the ballistic particle model. The
  cone (opening angle $\theta$) has walls of finite angular width ($b$).
  The cone, which is symmetrical about the binary line-of-centres, is
  rotated about the origin at an angle $i$ with respect to the
  observers line of sight (indicated by the bold solid line).}
\label{geom_fig}
\end{figure}

We constructed the dust density distribution using a simple ballistic
particle model in which mass packets are ejected from the conical
surface at the wind speed, and then follow linear trajectories (see
Figure~\ref{geom_fig}). The conical surface is rotated according to
the 245\,d period. This process leads to a spiral density structure.
The ensemble of particles is then read into {\sc torus}, and the
adaptive mesh is constructed.

\begin{figure}
\includegraphics[width=88mm]{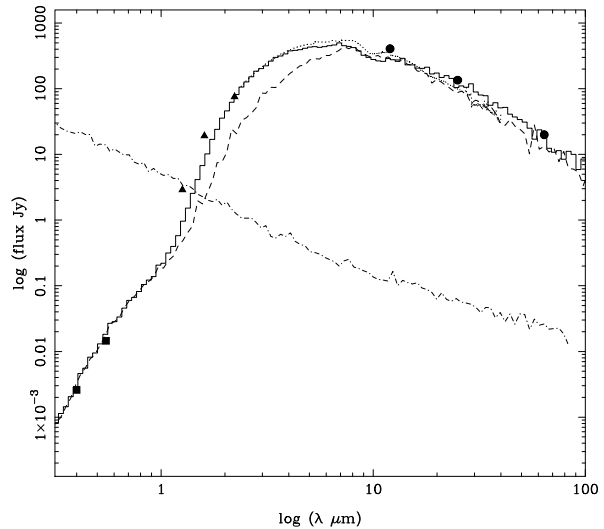}
\caption{Observed and model SEDs for
  WR104. The observations are the $B$ and $V$ photometry (solid
  squares), the $JHK$ data from Williams et al. (1987), the ISO-SWS
  spectrum (dotted line), and three IRAS data points (solid circles).
  The reference {\sc torus} SED is shown as a histogram, while the
  dot-dashed line indicates the model SED of the WR+O
  photospheres. The SED of the 1\micron\ grain model is also shown for
  comparison (dashed line).}
\label{sed_plot}
\end{figure}

\subsection{Free parameters}

We have fixed the binary parameters and distance, as well as the dust
geometry, chemistry and size distribution. We are then left with
effectively three free parameters:
\begin{itemize}
\item The dust mass. We compute spiral structure out to 100 binary
  revolutions (the level of the IR excess at 60\micron\ indicates
  that their is a significant amount of cool dust). The dust mass is
  then taken as the total mass in the spiral structure, and can be
  simply converted to a dust formation rate.
  
  The dust mass is constrained by the mid-to-far infrared (optically
  thin) parts of the SED. Furthermore, the mass must be great enough
  to ensure that optical depth in the first spiral revolution is
  sufficient to shield the outer turns from strong heating by the
  binary (constrained by the images).
  
\item The dust sublimation temperature. When the radiative equilibrium
  is calculated, some regions of the spiral will have a temperature
  greater than that at which dust can survive ($T_{\rm
    sub}$). After calculating the radiative equilibrium, cells whose
  temperature exceeds $T_{\rm sub}$ have their densities set to zero,
  and are excluded from the rest of the calculation. The density of
  the dust nebula is then rescaled to ensure that the input dust mass
  is conserved. A further radiative equilibrium iteration is then
  necessary (as parts of the spiral which were previously shielded
  from stellar radiation are now heated). This cycle of radiative
  equilibrium and cell rejection is repeated until no cells are
  rejected and temperature convergence is achieved. This process
  typically requires three or four iterations.
  
  The sublimation temperature is constrained by the near-to-mid IR
  section of the SED, since the peak of the SED is a measure of the
  dust temperature close to the central object.  If this value is too
  low, the dust in the inner spiral turn is destroyed, and the outer
  spiral becomes directly heated by the central binary, in
  contradiction with the images.
  
\item The foreground extinction. WR104 (along with WR105) is situated
  behind a strongly obscuring cloud (Lundstrom \& Stenholm 1984), and
  shows a total line-of-sight extinction of $A_V \approx 6.5^{\rm m}$ (e.g.
  Williams et al.  1987; Pendleton et al. 1994). It is also apparent
  that a significant amount of extinction is local to the system. Kato
  et al. (2002) showed that WR104 varies by over $2.5^{\rm m}$ in the
  optical, with marginal evidence for a periodicity of 245\,d in the
  light curve. Crowther (1997) argued that the variable obscuration
  arises from dust formation very close to the binary system. 
  
  The face-on spiral geometry employed here does not provide a
  mechanism for variable line-of-sight extinction.  It is possible
  that the variability results from the binary motion behind a sharp
  density gradient produced during a previous dust-producing episode
  prior to the current evolutionary phase. We therefore assume that
  all the extinction is occurring in a region external to the dusty
  spiral. We find that the 9.7\micron\ silicate feature seen in the
  ISO spectrum, which must be formed by interstellar grains, requires
  an optical depth of 0.3 of astronomical silicate (Draine \& Lee
  1984) to reproduce, and we fix this component of the optical depth
  in our reference model. The bulk of the extinction we ascribe to
  local material surrounding the system, but not associated with the
  current phase of dust production. Given the uncertain nature of the
  carbonaceous grains in the nebula, we adopt an MRN size distribution
  for the obscuring dust. We quantify the extinction by the optical
  depth of this dust at 5500\AA\ ($\tau_{5500}$).

  We emphasize that the nature of the foreground extinction has little
  impact on the SED beyond $\sim 1.5$\micron, and therefore does not
  strongly affect the other free parameters of our model, which are
  well constrained by the peak and long-wavelength tail of the SED.

\end{itemize}

In addition to the three parameters above, we also investigate the
changes to the SED and images from changing the opening-angle of the
conical wind-wind collision interface.

\begin{figure}
\includegraphics{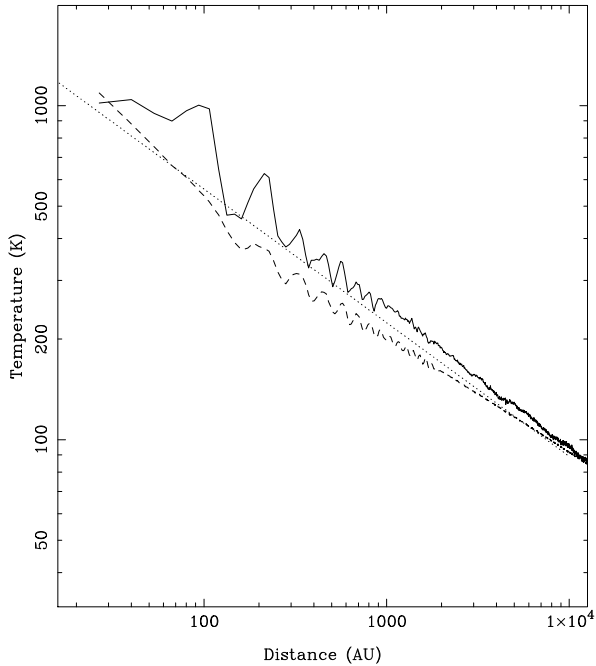}
\caption{Azimuthally-averaged temperature as a function of radial
  distance from the central binary. The reference model temperature
  structure (solid line) is plotted along with the model with
  1\micron\ grains (dashed line). The predicted curve for optically
  thin ($T \propto r^{-0.4}$) dust is shown in comparison to the
  reference model.}
\label{temp_plot}
\end{figure}

\section{The reference model}

A small grid of SEDs and images was constructed, varying the dust
formation rate, the dust sublimation temperature ($T_{\rm sub}$) and
the foreground optical depth. It was found that the best fit to the
SED had a dust formation rate of $8\pm 1 \times
10^{-7}$\,\msol\,yr$^{-1}$, $T_{\rm sub}=1200\pm 50$\,K, and $\tau_{\rm
  5500} = 3\pm 0.5$, and this set of parameters was adopted as our reference
model (see Table~\ref{ref_tab}).

\begin{table}
\caption{Basic parameters of the reference model.}
\label{ref_tab}
\begin{center}
\begin{tabular}{lll}
\hline
Parameter & Value  & Adopted/Fitted\\
\hline
Period (d) & 245 & Adopted\\
Inclination ($^\circ$) & 11 & Adopted\\
\mdot$_{\rm WR}$ (\msol\,yr$^{-1}$) & $3 \times 10^{-5}$ & Adopted\\
\mdot$_{\rm O}$ (\msol\,yr$^{-1}$)y & $6 \times 10^{-8}$ & Adopted\\
$v_{\infty,{\rm WR}}$ (\kms) & 1220 & Adopted \\
$v_{\infty,{\rm O}}$ (\kms) & 2000 & Adopted \\
Cone angle ($^\circ$) & 40 & Adopted \\
$T_{\rm sub}$ (K) & 1200 & Fitted \\
Dust speed (\kms) & 845 & Adopted\\
\mdot$_{\rm dust}$ (\msol\,yr$^{-1}$) & $8 \times 10^{-7}$ & Fitted\\
Distance (kpc) & 1.6 & Adopted\\
\hline
\end{tabular}
\end{center}
\end{table}

The outer turns of the spiral are optically thin, and the temperature
structure in this region is  expected to be
\begin{equation}
T(r) \propto r^{-2/(4+p)}
\end{equation}
where $\kappa_\lambda \propto \lambda^{-p}$. The small grains have an
opacity distribution that goes as $\lambda^{-1}$, and we therefore
expect $T(r) \propto r^{-0.4}$ at large distances (Spitzer 1978).

\begin{figure}
\includegraphics[height=88mm]{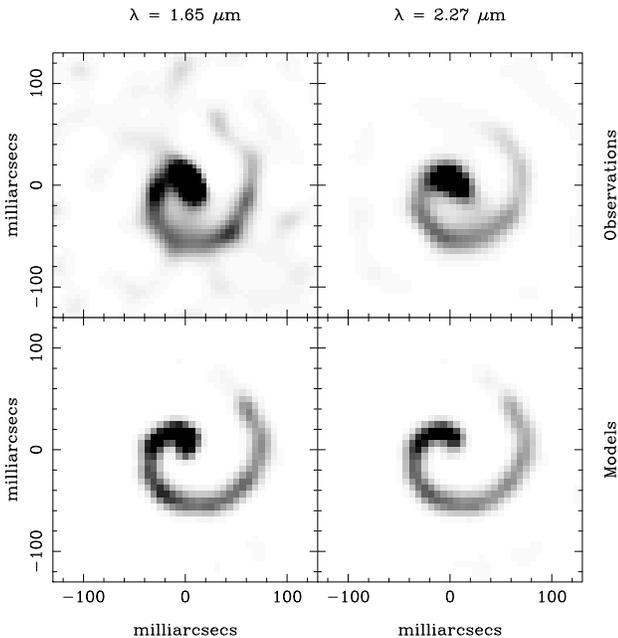}
\caption{A comparison between the observed (top panels) and model (top
  panels) NIR images of WR104. The $H$ (1.65\micron) and CH4
  (2.27\micron) observations are shown in the top panels, and the
  synthetic images (computed using the same filter bandpasses) are
  shown underneath. The greyscale shows intensity on a linear scale
  starting at zero (white), and constructed so that mid-grey
  corresponds to the intensity at the southernmost turn of each
  spiral.}
\label{images_plot}
\end{figure}

In Figure~\ref{temp_plot} we plot the azimuthally-averaged radial
temperature distribution of the reference model, and the outer turns
of the spiral do indeed approximately follow a $T(r) \propto r^{-0.4}$
relation. There is a low temperature gradient along the first turn of
the spiral, since much of the dust is very close to $T_{\rm sub}$. The
subsequent turn of the spiral, however, shows a significant
temperature drop, since it is effectively shielded from the binary
radiation. A transition between the optically thick and optically thin
temperature gradients occurs at approximately the third spiral turn,
and beyond here the $T(r) \propto r^{-0.4}$ applies.

The model SED (Figure~\ref{sed_plot}) shows reasonable agreement with
the data, but once again we stress that the observations in the
various wavelength regimes are not contemporaneous. The radiation from
the spiral is dominated by thermal dust emission, with a negligible
scattered component due to the small grain albedo and the fact that
the emission occurs far into the Rayleigh-Jeans tail of the combined
WR+O SED.

In Figure~\ref{images_plot} we show greyscale images of the
1.65\micron\ and 2.27\micron\ models and the data. In order to make a
direct comparison we have used the same pixel scale of the
observations for our model images, and have smoothed them by a 2-pixel
(12\,mas) FWHM Gaussian to simulate the finite resolution of the Keck
masking interferometry. We have scaled the images in order that the
greyscales match in the southernmost part of the spiral, since the
peak intensity of the observation is compromised by artifacts
introduced by maximum entropy image construction.  Qualitatively the
model images show reasonable agreement with the interferometry, in
particular the extent of the spiral is well-matched, and the increase
in flux towards the centre of the spiral. It is interesting to note
that the `head' of the spiral 1.65\micron\ is slightly more inclined
than that of the 2.27\micron\ image, a pattern that is reproduced in
the models, since there is significant photospheric flux at
1.65\micron.
\begin{figure}
\includegraphics[height=88mm,angle=90]{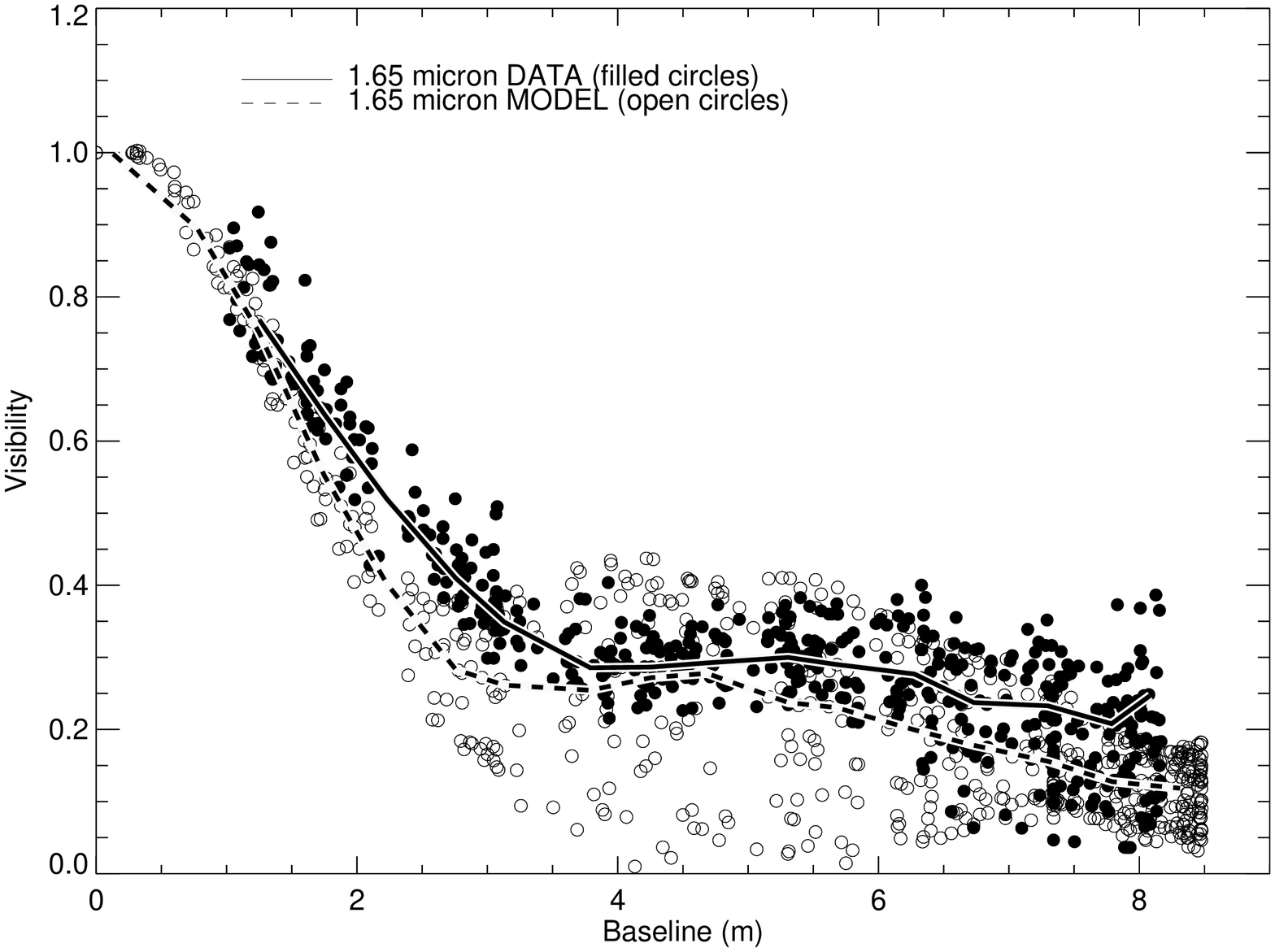}
\includegraphics[height=88mm,angle=90]{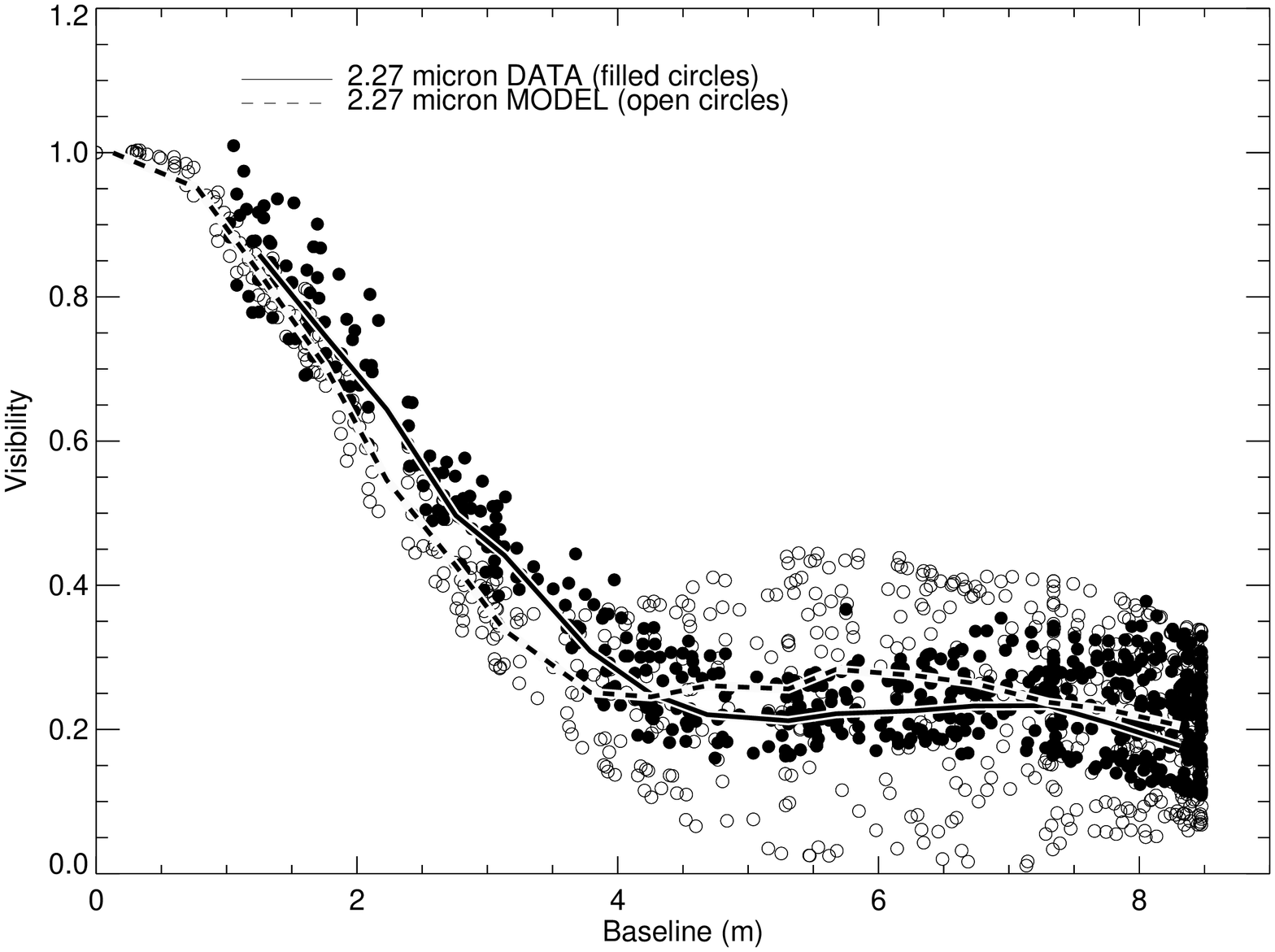}
\caption{Visibility curves for the 1.65\micron\ (upper panel) and
  2.27\micron\ (lower panel) passbands. The observational data (filled
  circles) and compared to the models (open circles, dashed lines).
  The solid lines are azimuthal averages of the data.}
\label{viscurve_plot}
\end{figure}

Figure~\ref{viscurve_plot} shows a more quantitative comparison
between the 1998 April model images and the measured visibility data
from aperture masking.  In creating this figure, we sampled the model
images at the same $(u,v)$ points as were measured by the aperture mask.
The large number of points arises because we sample the visibility
data using a 36-hole `pseudo'-array (630 baselines) which
approximates the continuous Fourier coverage afforded by the `annulus'
aperture mask. Note that data points below 1-m baselines were omitted
due to poor calibration within the `seeing spike' (see Monnier et al.
2004 for an extensive discussion of calibration issues of Keck
aperture masking).

Given the strong resemblance of the model images and the MEM
reconstructions in Figure~\ref{images_plot}, the overall good agreement
of the visibility data with the model comes as no surprise.  In
particular, the 1.65 and 2.27~micron visibility data agree well with
model when we consider the azimuthally-averaged behavior (especially
in light that the model parameters were not optimized to fit this
depiction of the data). That is to say that the model has the correct
characteristic sizes at these wavelengths and the right balance of
compact features (unresolved on the longest baselines) and diffuse
structure (resolved on the longest baselines).  However, the model
visibilities show much larger variation as a function of position
angle for a given baseline length (especially at intermediate
baselines between 4--7\,m).  Simple changes of scale or spiral width do
not change this discrepancy, and we hypothesize that this difference
arises due to the difficulties in modelling the shape and relative
brightness of the central dust condensation.

\subsection{Deviations from the reference model}

In this section we attempt to quantify the changes in the model that
occur if we alter the parameters of the reference model.

\subsection{Inclination}
\label{inc_sec}

\begin{figure}
\includegraphics[width=88mm]{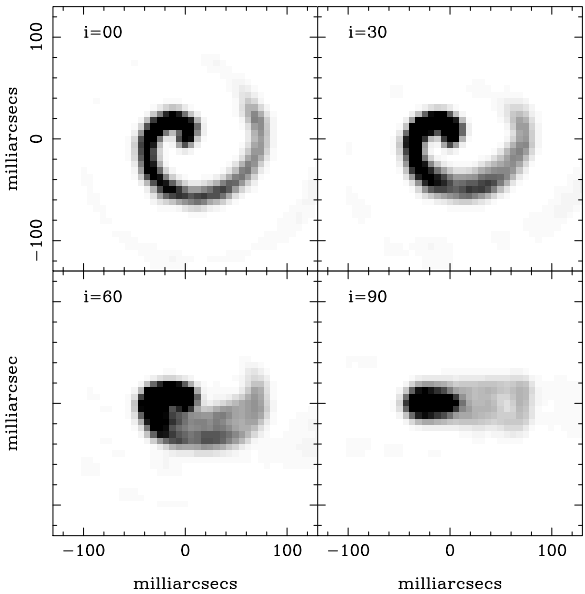}
\caption{Model H-band images viewed at various inclinations.}
\label{inc_plot}
\end{figure}

We fixed the inclination at 11$^\circ$ in the reference model, a value
that was determined by Tuthill et al. (1999) by fitting a linear
spiral model to the Keck data. We have investigated the effect on the
projected images of changing this value, computing images at
inclinations of 0, 30, 60 and 90$^\circ$ (Figure~\ref{inc_plot}). The
spiral nature of the spiral is easily discernible in the $i=30^\circ$
image, although such a high inclination is ruled out by the
interferometry. 

Inevitably the higher ($i>30^\circ$) inclination images show much less
evidence for the spiral nature of the nebula, although such images
will be useful for testing the spiral model against systems that are
seen closer to edge on, such as WR137 (Marchenko et al. 1999) and
WR140 (Monnier et al. 2002).

\subsubsection{Cone angle}
\label{angle_sec}

The cone angle of the wind-wind collision surface depends on the ratio
of the wind momenta of the binary components
(equation~\ref{theta_eq}).  Since the imaging data may provide a
constraint on this quantity we have computed further models, based on
the reference parameters but with a semi-opening angle of 80$^\circ$
and $120^\circ$ (we retain the 5$^\circ$ spread in the flow).

\begin{figure*}
\includegraphics[width=188mm]{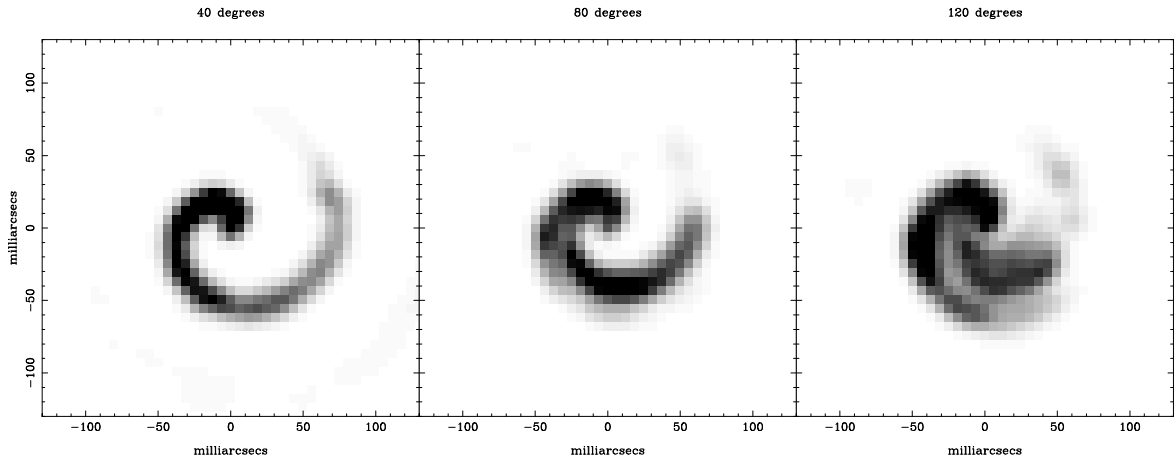}
\caption{Results of varying the opening angle of the wind-wind
  collision interface. The left-hand panel shows the reference model
  (opening angle is $40^\circ$), the middle panel has an opening angle
  of $80^\circ$, the right-hand panel was computed with an opening
  angle of $120^\circ$.}
\label{cone_plot}
\end{figure*}

Figure~\ref{cone_plot} illustrates the effect of varying the cone
angle while holding the dust production rate constant leads to a
lower-density structure, although the inner turn of the spiral remains
optically thick.  A larger surface of the inner spiral is heated, and
as such the 1--10\micron\ flux from the model increases (the long
wavelength flux, where the dust is optically thin, is unaffected by
the change in geometry). Increasing the cone angle to 120$^\circ$
leads to a pronounced change in the near-IR image
(Figure~\ref{cone_plot}), and such a wide angle is clearly ruled out by
the interferometry. The visibility curves for the $80^\circ$ model
show that the nebula is too resolved (the visibility drops too fast at
short baselines), and that the visibility is too low at long
baselines, indicating that the model lacks a central
concentration. 

\subsubsection{Grain size}
\label{grainsize_sec}
A uniform grain size of 0.01\micron\ was used for the reference model,
based on the theoretic predictions of Zubko (1998). However, there is
evidence that some of the grains in the nebula might be considerably
larger than this: Chiar \& Tielens (2001) found that a
grain size of 1\micron\ was required to fit the 6.2\micron\ absorption
profile in ISO spectra of WC stars (including WR104), and Marchenko et
al. (2002) argued that similarly large (0.5\micron) grains were
necessary to match the extended dust emission from WR112.

In order to investigate the effect of grain size on our models we
computed images and SEDs for a model with a uniform grain size of
1\micron. The opacity distribution for these grains is shown in
Figure~\ref{albedo_fig}. It can be seen that in the long wavelength
(Rayleigh) regime the opacity is independent of grain size (and so
changes in the dust production rate affect both the small and large
grain models equally). The large grains are effectively grey
shortwards of 10\micron, and the albedo of the dust is much higher
($\sim 0.5$) than the 100\AA\ grains for $\lambda<100$\micron.
However the opacity of the large grains is two orders of magntiude
lower than the small grains near the peak of the binary flux
distribution.

The temperature structure of the large grain model is flatter than
that of the reference model, with an $r^{-0.4}$ dependence across
virtually the entire computational domain (see
Figure~\ref{temp_plot}). Close to the central stars the temperature
gradient is marginally steeper, corresponding to a regime where the
spiral turns are getting optically thick.

\begin{figure}
\includegraphics[width=88mm]{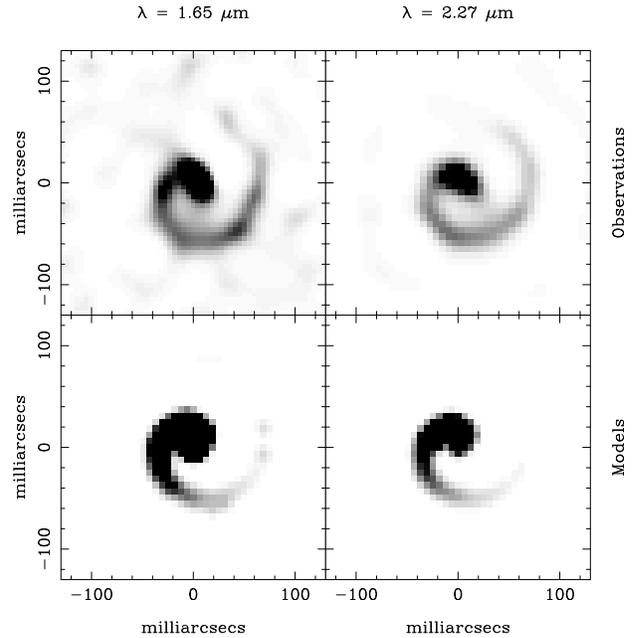}
\caption{Model images assuming a uniform grain size of 1\micron.}
\label{biggrain_plot}
\end{figure}

The model images are significantly different to those of the reference
model (see Figure~\ref{biggrain_plot}). For the same dust sublimation
temperature the large-grain dust may exist much closer to the central
stars, and therefore the emission is much more centrally condensed.
The more rapid decline in temperature along the first spiral manifests
itself as a sharp intensity gradient along the innermost spiral in
both synthetic images. Naturally the SED of the large grain model
shows good agreement with the reference model at short wavelengths
(where photospheric emission is dominant) and at long wavelengths
(where the grain opacity is the same and the dust optically thin). The
SEDs deviate in the near-to-mid IR ($\sim 2$--10\micron), with the
large grain model showing reduced emission in this region
(Figure~\ref{sed_plot}). This discrepancy is due to the smaller mass
of dust near the dust sublimation temperature in the large grain
model. Increasing the dust sublimation temperature does not
sufficiently increase the flux in this part of the SED, while
increasing the dust production rate degrades the agreement with the
observations at long wavelengths.

\section{Discussion and conclusions}

The reference model has a dust production rate of $8 \pm 1 \times
10^{-7}$\,\msol\,yr$^{-1}$, which compares reasonably well with that
derived by Zubko (1998), who used a spherical shell model and found
$5.3 \times 10^{-7}$\,\msol\,yr$^{-1}$. Similarly, we find a dust
sublimation temperature of 1200\,K, which is within 20\,K of Zubko's
value.  The agreement is perhaps unsurprising -- the leverage on both
quantities comes from the SED, and the mass-loss rate itself is
relatively insensitive to the adopted geometry. However, Zubko also
noted that the dust in his WR104 model was optically thick in the UV,
which rendered his model `an approximation'. We also find an optically
thick geometry, with $\tau_{5500} \sim 3$, across the densest part of
the spiral, although our radiative equilibrium method properly treats
these dense regions. If we take a mass-loss rate of $3 \times 10^{-5}$
\,\msol\,yr$^{-1}$ for the WR star (Crowther 1997), and a carbon
mass-fraction of 0.5 (Crowther 1997), we find that approximately 2\%
of the available carbon in the wind condenses into dust.

The dust production rate derived from the reference model is
insensitive to the dust grain size distribution, since it is primarily
constrained by the long wavelength parts of the SED, where the the
dust opacity is independent of size. However the combination of
spectral {\em and} imaging data allow us to constrain the dust grain
size using the near-to-mid IR region of the SED. We find that for the
small dust model the high opacity to UV radiation, and the proximity
of the dust to the central object determined from the imaging, means
that the inner-edge of the first turn of the spiral is heated to close
to the dust sublimation temperature. Since the opacity of the large
grain model is much (two orders of magnitude) lower in the UV, the
heating correspondingly less and we see a smooth $r^{-0.4}$ temperature
gradient along the inner spiral turn, with only the tip of the spiral
at the dust sublimation temperature. We therefore prefer the small
grain model, although we cannot rule out larger grains further out in
the nebula. It is interesting to note that Marchenko et al. (2003)
found that small grains ($\sim700$\AA) were required to explain the
optical extinction to WR140 during periastron passage. The inescapable
conclusion is that the dust nebulae around WC binaries contain a range
of dust sizes from 0.01--1\micron. The presence of such large grains
would require modification to grain condensation models, which predict
rather slow growth once the grains reach 100\AA\ (Zubko 1998). It is
possible that incorporating a more realistic geometry in the dust
chemistry calculations, such as the spiral presented here, will allow
significant grain growth since large parts of the nebula are shielded
from direct stellar radiation and yet still have a relatively high
density.

The interferometric imaging appears to rule out opening angles for the
wind-wind collision interface of $\ge 80^\circ$, and the opening angle
($40^\circ$) adopted in the reference model provides a good fit to
both the SED and the images, which lends further support to the binary
component parameters estimated by Crowther (1997). Our estimate for
the cone angle is similar to that found for WR137 ($40$--$60^\circ$;
Marchenko et al. 1999) and is slightly smaller than that of WR140,
which has estimates of $\sim 70^\circ$ (Monnier, Tuthill \& Danchi
2002) and $\sim 80^\circ$ (Marchenko et al. 2003). 

Overall the models seem to support the basic paradigm for WR104 of a
linear archimedean spiral produced by the rotation of a conical
wind-wind collision interface. There are still deficiencies in the
model, in particular regarding the structure and opacity of the
material very close to the binary, where a more sophisticated
description of the geometry may be necessary. For example, we have so
far only considered dust opacity, but there will be significant gas
opacity in the intra-binary region, which will play a role in
reprocessing the direct stellar radiation in the innermost parts of
the nebula. Despite these shortcomings, the model is very successful
in predicting the gross observational properties of WR104, and also
enables us to constrain the dust properties.

WR104 represents perhaps the simplest of the resolved WC+O dust
nebulae to interpret due to its face-on orientation and the apparent
circularity of its orbit. Nebulae with much more complicated
structures occur when the binary is viewed at a higher inclination or
if the orbit is eccentric (for example WR98a, WR112, WR137 and WR140).
Three-dimensional radiative-transfer modelling represents the only
route to deconvolving the true nebula geometry from the projected
images in these more complex cases, and further modelling of other
colliding wind systems with resolved dust nebulae should provide
detailed insights into dust formation in the harsh environments of
massive binary systems.

\section*{Acknowledgements}

We are grateful to Leon Lucy for advice on implementing his radiative
equilibrium method. Ilaria Pascucci is warmly thanked for providing
the 2-D benchmark SEDs and temperature runs. Paul Crowther is thanked
for providing the CMFGEN model SED of WR104. We thank the authors for
providing the radiative-transfer code {\sc dusty}, which is available
from {\tt http://www.pa.uky.edu/$\sim$moshe/dusty/}. This research
made use of the ISO Data Archive, the SIMBAD database, and the ADS
abstract service.  RK is funded by PPARC standard grant
PPA/G/S/2001/00081. Some of the data herein were obtained at the W.M.
Keck Observatory, made possible by the support of the W.M. Keck
Foundation.

\end{document}